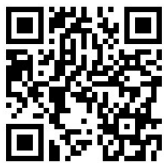

## NOTAS Y EXPERIENCIAS / *NOTES AND EXPERIENCES*

# Analysis of the coverage of the Data Citation Index – Thomson Reuters: disciplines, document types and repositories

Daniel Torres-Salinas*, Alberto Martín-Martín**, Enrique Fuente-Gutiérrez***

*Grupo Evaluación de la Ciencia y la Comunicación Científica, Centro para la Investigación Médica Aplicada, Universidad de Navarra.
Correo-e:torressalinas@gmail.com
**Grupo Evaluación de la Ciencia y la Comunicación Científica, Facultad de Comunicación y Documentación. Universidad de Granada.
Correo-e: albertomartin101@gmail.com
*** Grupo Evaluación de la Ciencia y la Comunicación Científica, Facultad de Comunicación y Documentación. Universidad de Granada.
Correo-e: enfun02@gmail.com

**Abstract:** In the past years, the movement of data sharing has been enjoying great popularity. Within this context, Thomson Reuters launched at the end of 2012 a new product inside the Web of Knowledge family: the Data Citation Index. The aim of this new database is to enable discovery and access, from a single place, to data from a variety of data repositories from different subject areas and from around the world. In short note we present some results from the analysis of the Data Citation Index. Specifically, we address the following issues: discipline coverage, data types present in the database and repositories that were included at the time of the study.

**Análisis de la cobertura del Data Citation Index – Thomson Reuters: disciplinas, tipologías documentales y repositorios**

**Resumen:** En los últimos años, el movimiento conocido como "data sharing", es decir compartir lo datos de investigación, está cobrando una gran popularidad. Dentro de este contexto Thomson Reuters lanzó a finales de 2012 un nuevo producto dentro de su plataforma Web of Knowledge: el Data Citation Index. El objetivo de esta nueva base de datos es facilitar el acceso desde un único punto a los datos indexados en diferentes repositorios de datos de todo el mundo. En esta nota se presentan los resultados del análisis del Data Citation Index y más concretamente se analiza la cobertura de este producto atendiendo a las disciplinas, las tipologías documentales indexadas y los repositorios que se encuentran disponibles en el momento de la realización del estudio.

**Palabras clave:** Datos; datos de investigación; acceso abierto; data sharing; comunicación científica; bases de datos; indices de citas; Thomson Reuters

## 1. INTRODUCTION

During the last decade, there has been a heated debate among the scientific community about the need of releasing research data, a movement commonly referred to as data sharing. Although the practice of sharing data has been present among researchers for a long time (Hrynaszkiewicz, Altman, 2009), the movement of data sharing is currently enjoying great popularity due to the convergence of a number of circumstances, two of the most important being the development of the information technologies, and researcher's ever more open attitude towards their findings (as exemplified by movements like Open Access).

The benefits of data sharing have already been studied and identified (Arzberger et al., 2004; Vickers, 2006). In the first place, data sharing contributes to make the most of the funds invested in science because it helps prevent duplication of efforts and also because it makes possible the development of new studies that reuse these data. This is worth considering in the present situation of economic crisis, especially when research is government funded. Secondly, these data can be used as a tool to detect fraud, since they would enable other researchers to verify or disprove the results of an experiment through its replication (Renolls, 1997). Thirdly, there is evidence that published studies whose data are openly available receive more citations (Piwowar, Day, Fridsma, 2007). Lastly, it is possible that these practices open the way for the creation of data metrics that complement existing indicators for scientific evaluation (Wouters and Schröder, 2003; Costas et al., 2013).

Currently there are a large number of initiatives, commonly called data banks or data repositories, dedicated to store, describe and disseminate scientific data. Unlike pre-prints or post-prints repositories, which deal only with one bibliographic format for the items they contain, there is a great variety of data repositories and the solutions adopted are different in each case, and often this makes them difficult to use to people without knowledge of the data bank's subject area (Torres-Salinas et al., 2012).

Within the context described above, Thomson Reuters has added a new member to the Web of Knowledge family of databases: the Data Citation Index (henceforth DCI). The DCI, released in November 2012, is described as a tool to discover and access, from a single place, data from a variety of repositories from the three major subject areas (Science & Technology, Social Sciences, and Arts & Humanities) and from around the world. In order to be included in the DCI, a data repository must first undergo a process of evaluation in which a number of factors are considered, including the repository's basic publishing standards, its editorial content, the international diversity of its authorship, and the citation data associated with it (Thomson Reuters, 2012). At the same time, records in the DCI are linked to the publications they inform, thus providing citation information for the data sets, and opening the way to data citation analysis. However, even though the DCI is the first tool that allows us to quantify the impact and reutilization of research data, it is as of yet a young product that needs to be assessed in order to comprehend its strengths and limitations. This assessment will allow bibliometricians, librarians, and the rest of potential users of this tool to better understand for what purposes it may be used and how.

For this reason in this note we present an analysis of this new database; more specifically we address the following questions:

Question 1. What is the discipline and subject area coverage in the DCI?

Question 2. What kinds of data types are present in the DCI, and what is their statistical distribution?

Question 3. Which repositories contribute a larger share of records to the DCI and what are their basic characteristics (data type, country, etc..)?

These results are interesting since they are the first empiric results obtained from an analysis of the DCI as a scientific information and evaluation tool. We should also mention that this note is based on a previous working paper deposited in Arxiv in June 2013 (Torres-Salinas et al., 2013).

## 2. METHODOLOGY

For the purpose of this analysis, all records from the Data Citation Index were downloaded in April-May 2013, using the DCI web interface. The resulting text files were processed and added to a relational database, using the Accession Number Field (UT) as the primary key for the data records. The rest of the fields analyzed were: Document Type (DT), Publication Year (PY), and Web of Science Category (WC). Regarding the issue of discipline coverage, two classification systems have been used in order to assign categories to the records: one of them comprises four major subject areas (Science, Social Sciences, Humanities & Arts, and Engineering & Technology), and the other is the one proposed by Moed (2005), with thirteen disciplines. These systems were built by aggregating Web of Science categories, in the same way as we did in other studies analyzing products by Thomson Reuters (Torres-Salinas et al., 2013).

## 3. RESULTS

### 3.1. General description and distribution per area and scientific field

At the time of the download, the Data Citation Index held a total of 2.623.528 records. The oldest of them can be traced back to the year 1800 (Figure 1) but, as expected, this database mainly deals with contemporary data, and 92% of records are dated between 2000 and 2013. The year where we can find more records is 2009, with a total of 365,381. If we attend to their subject areas, it is clear that most of the records belong to the area of Science, with a crushing 80% (Figure 2), well ahead of the Social Sciences with 18%, and Humanities & Arts with 2%. The presence of records in the area of Engineering & Technology is almost non-existent, with less than 0.1%. These results are consistent with the known issue of the under-representation of the Social Sciences and Arts & Humanities in other multidisciplinary databases of the Web of Science family.

If we consider the classification system proposed by Moed (Figure 2), Clinical Medicine is the discipline that accounts for the largest share of the records (50.86%), closely followed by Molecular Biology and Biochemistry with 47.96%, and, at some distance, Geosciences with 20.12% (note that a record may be assigned to several disciplines).

**Figure 1.** Record distribution in the Data Citation Index by year of publication

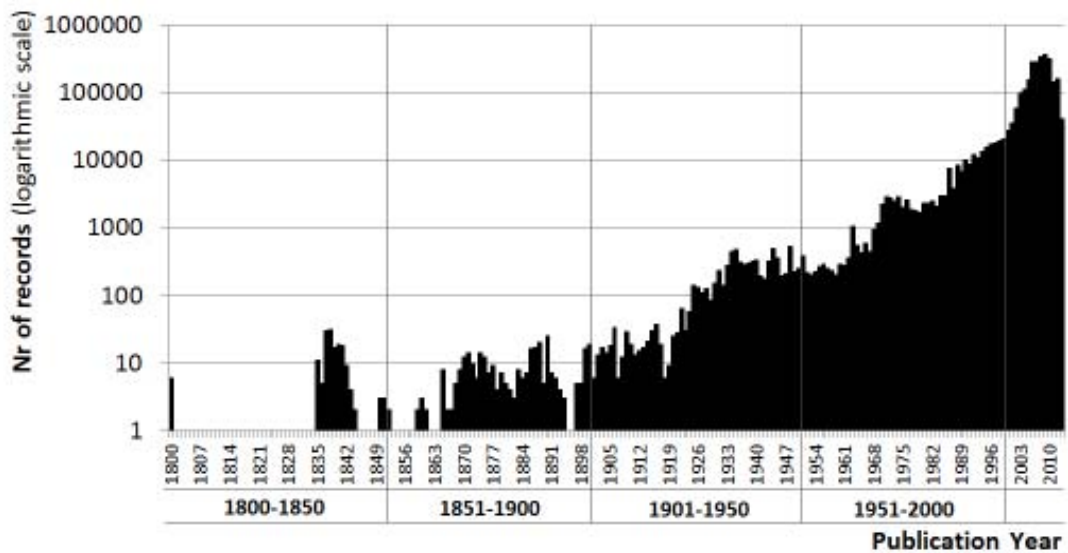

**Figure 2.** Record distribution in the Data Citation per discipline and scientific field

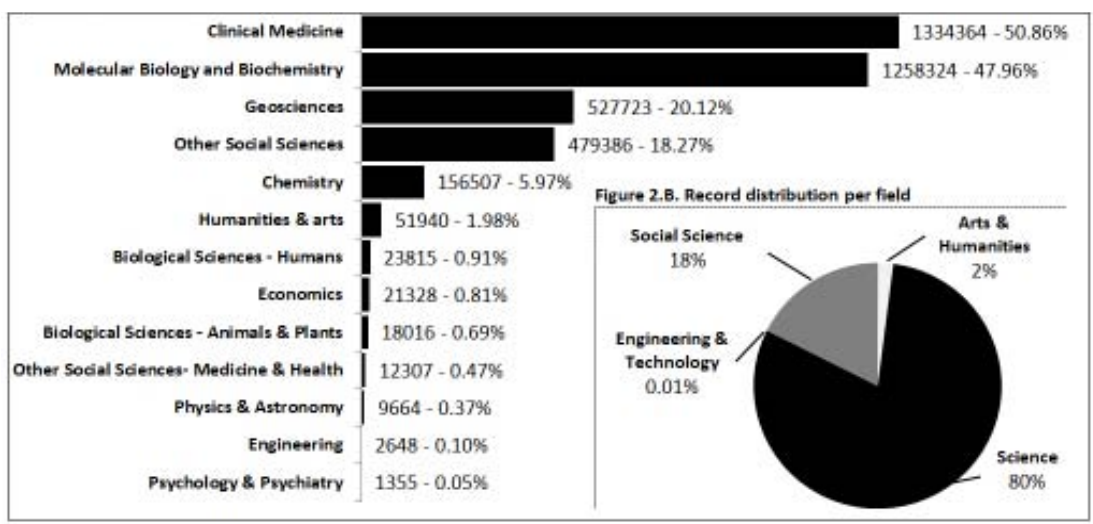

### 3.2. Distribution per type of document

The Data Citation Index contains at the moment three different document types: data repositories, data studies, and data sets (Thomson Reuters, 2012). Data sets are the basic unit of information and are usually, but not necessarily, part of a data study. Thomson Reuters define data set as a single or coherent set of data or a data file provided by the repository, as part of a collection, data study or experiment. As for data studies they are, according to Thomson Reuters, description of studies or experiments held in repositories with the associated data which have been used in the data study. The distribution of all records among each of these document types is presented in Table I, broken down by subject areas. There are a total of 2,475,534 records in the data set category, which makes it the most common document type in the database by far, with 94% of the total number of records. Only 159,280 are classified as a data study (6%) and 96 as a data repository. As shown in Table I, Science accumulates 81% of all the data sets and 73.92% of data studies. Data sets are also the predominant typology in every major subject area. It is also worth noticing that there seems to be a larger presence of data studies in the areas of Engineering & Technology, and Humanities & Arts (around 13% of the total of records in both areas) which doubles the average percentage for that document type if we consider the entire database (6%).

### 3.3. Main repositories and distribution

Lastly, in Table II we present the names and record count of the main repositories that are indexed in the DCI. We only consider those repositories which contain at least 4000 records, regardless of the document type. Only 29 repositories met this requirement. Also we found 64 repositories that contain at least 100 records. As can be seen, there is a very high concentration of records in a set of four repositories, which account for 75% of records in the DCI: Gene Expression Omnibus, UniProt Knowledgebase, PANGAEA and U.S. Census Bureau TIGER/Line Shapefiles. Regarding disciplines the first two repositories belong to Biochemistry & Molecular Biology, and Genetics & Heredity, while the other two fall within the scope of Geosciences, Social Sciences, and Geography. The best represented disciplines in the DCI in terms of number of repositories are Genetics & Heredity (24), Biochemistry & Molecular Biology (16), Social Sciences, Interdisciplinary (13), Astronomy & Astrophysics (9) and Geosciences, Multidisciplinary (9). Other interesting aspects are that most of the repositories in Table II are specialized in data sets, there exists a predominance of repositories managed by universities (45%) and most of them are located in the United States (59%).

**Table I.** Document type distribution by field in the Data Citation Index

| Table I.A. Number of records per field and type of document | | | | |
|---|---|---|---|---|
| | Data set | Data study | Repository | **Total** |
| Engineering & Technology | 1,545 | 240 | 1 | **1,786** |
| Humanities & Arts | 44,588 | 6,847 | 9 | **51,444** |
| Science | 2,004,449 | 114,338 | 67 | **2118,855** |
| Social Sciences | 424,952 | 37,855 | 19 | **462,826** |
| **Total** | **2,475,534** | **159,280** | **96** | **263,4911** |
| **Table I.B. Percentage of records per field and type of document** | | | | |
| | Data set | Data study | Repository | **Total** |
| Engineering & Technology | 86.51% | 13.44% | 0.06% | **100%** |
| Humanities & Arts | 86.67% | 13.31% | 0.02% | **100%** |
| Science | 94.60% | 5.40% | 0.00% | **100%** |
| Social Sciences | 91.82% | 8.18% | 0.00% | **100%** |
| **Total** | **93.95%** | **6.04%** | **0.00%** | **100%** |
| **Table I.C. Percentage of records per type of document and field** | | | | |
| | Data set | Data study | Repository | **Total** |
| Engineering & Technology | 0.06% | 0.16% | 1.11% | **0.07%** |
| Humanities & Arts | 1.81% | 4.43% | 10.00% | **1.96%** |
| Science | 81.19% | 73.92% | 75.56% | **80.76%** |
| Social Sciences | 17.21% | 24.47% | 21.11% | **17.64%** |
| **Total** | **100%** | **100%** | **100%** | **100%** |

Table II. Main repositories in the Data Citation Index sorted by number of records

| Repository Denomination | Nr of Records in DCI | % From the total DCI | % Accum. From the total DCI | % Dataset In DCI | % Data Study In DCI | Web of Science Category | Country/Region | Type of Institution Responsible |
|---|---|---|---|---|---|---|---|---|
| Gene Expression Omnibus | 654,917 | 24.96% | 24.96% | 97% | 3% | Biochemistry & Molecular Biology \| Genetics | USA | Research Center |
| UniProt Knowledgebase | 496,803 | 18.94% | 43.90% | 100% | --- | Biochemistry & Molecular Biology \| Genetics | Multinational | Research Center |
| PANGAEA | 447,137 | 17.04% | 60.94% | 99% | 1% | Geosciences, Multidisciplinary | Germany | University |
| U.S. Census Bureau TIGER/Line Shapefiles | 358,957 | 13.68% | 74.63% | 100% | --- | Geography | USA | Government |
| Crystallography Open Database | 150,917 | 5.75% | 80.38% | 100% | --- | Crystallography | Multinational | Various |
| ArrayExpress Archive | 91,846 | 3.50% | 83.88% | 71% | 29% | Genetics & Heredity | Europe | Research Center |
| Protein Data Bank | 76,563 | 2.92% | 86.80% | 100% | --- | Biochemistry & Molecular Biology | Multinational | Various |
| Inter-university Consortium for Political and Social Research | 72,637 | 2.77% | 89.57% | 89% | 11% | Social Sciences, Interdisciplinary | USA | University |
| Roper Center for Public Opinion Research | 25,384 | 0.97% | 90.53% | 52% | 48% | Social Sciences, Interdisciplinary | USA | University |
| U.S. National Oceanographic Data Center | 25,370 | 0.97% | 91.50% | --- | 100% | Oceanography | USA | Government |
| EMAGE Gene Expression Database | 23,566 | 0.90% | 92.40% | 100% | --- | Genetics & Heredity | United Kingdom | Hospital |
| miRBase | 18,222 | 0.69% | 93.09% | 100% | --- | Genetics & Heredity | United Kingdom | University |
| Animal QTL Database | 16,636 | 0.63% | 93.73% | 100% | --- | Genetics & Heredity | USA | University |
| NOAA National Geophysical Data Center | 16,500 | 0.63% | 94.36% | 100% | --- | Geosciences & Geophysics | USA | Government |
| Institute for Quantitative Social Science | 16,196 | 0.62% | 94.97% | 64% | 36% | Social Sciences, Interdisciplinary | USA | Research Center |
| Odum Institute Data Archive | 10,516 | 0.40% | 95.38% | 55% | 45% | Social Sciences, Interdisciplinary | USA | University |
| IEDA: Marine Geoscience Data System | 9,110 | 0.35% | 95.72% | 81% | 19% | Geosciences, Multidisciplinary \| Oceanography | USA | University |
| nmrshiftdb2 | 8,962 | 0.34% | 96.06% | 54% | 46% | Physics, Atomic, Molecular & Chemical | Germany | University |
| Chemical Effects in Biological Systems | 8,939 | 0.34% | 96.40% | --- | 100% | Environmental Sciences | USA | Research Center |
| The Cell: An Image Library | 8,789 | 0.34% | 96.74% | 100% | --- | Cell Biology | USA | Various |
| Dryad | 6,639 | 0.25% | 96.99% | 68% | 32% | Ecology \| Evolutionary Biology \|Biodiversity | USA | University |
| NOAA Paleoclimatology | 6,522 | 0.25% | 97.24% | --- | 100% | Geosciences, Multidisciplinary | USA | Government |
| Cancer Models Database | 5,935 | 0.23% | 97.47% | 100% | --- | Genetics & Heredity | USA | Various |
| Nucleic Acid Database | 5,596 | 0.21% | 97.68% | 100% | --- | Biochemistry & Molecular Biology \| Genetics | USA | University |
| The Association of Religion Data Archives | 5,405 | 0.21% | 97.89% | 88% | 12% | Religion | USA | University |
| Eurostat | 5,366 | 0.20% | 98.09% | 93% | 7% | Social Sciences, Interdisciplinary | Europe | Government |
| UK Data Archive | 4,965 | 0.19% | 98.28% | --- | 100% | Social Sciences, Interdisciplinary | United Kingdom | University |
| DrugBank | 4,743 | 0.18% | 98.46% | --- | 100% | Pharmacology & Pharmacy | Canada | University |
| International Food Policy Research Institute | 4,351 | 0.17% | 98.63% | 98% | 2% | Food Science \| Ethnic Studies \|Demography | USA | Research Center |
| All the rest 61 repositories | 36,011 | 1.37% | 100.00% | --- | --- | ---- | ---- | |

## 4. CONCLUDING REMARKS

In this note we have presented some preliminary results based on the analysis of the Data Citation Index. We have shown discipline coverage, the data repositories and document types that can be found in this new database. The main conclusions and findings about the DCI can be summarized as follows:

1) It is heavily oriented towards the hard sciences; Science accounts for 80% of the records in the database. Within this area, the best represented disciplines are Clinical Medicine, Genetics & Heredity, and Biochemistry & Molecular Biology.

2) The DCI uses three document types (data set, data study and repository). There are 96 data repositories, and the predominant typology is the data set, with 2,475,534 records, which is 94% of the entire database.

3) Even though there are a total of 29 repositories that contain at least 4000 records, a total of 64 repositories that contain at least 100 records, there are four repositories that contain 75% of all the records in the database: Gene Expression Omnibus, UniProt Knowledgebase, PANGAEA, and U.S. Census Bureau TIGER/Line Shapefiles.

## 5. NOTES

This article is based on a previous working paper deposited in Arxiv in June 2013: Torres-Salinas, D.; Martín-Martín, A.; Fuente-Gutiérrez, E. (2013). An introduction to the coverage of the Data Citation Index (Thomson-Reuters): disciplines, document types and repositories. EC3 Working Papers (11), June 2013. http://arxiv.org/ftp/arxiv/papers/1306/1306.6584.pdf [Accessed on July 15 2013]

This article was written as part of the University of Granada´s "Introduction to Scientific Research" Grant Program.

This article has been translated by Alberto Martín-Martín and Nicolás Robinson-García.

## 6. REFERENCES

Arzberger, P.; Schroeder, P.; Beaulieu, A.; Bowker, G.; Casey, K.; Laaksonen, L.; Moorman, D.; Uhlir, P.; Wouters, P. (2004). An international framework to promote access to data. *Science*, vol. 303 (5665), 1777-1778. http://dx.doi.org/10.1126/science.1095958, PMid:15031482.

Costas, R., Meijer, I., Zahedi, Z. and Wouters, P. (2013). The Value of Research Data - Metrics for datasets from a cultural and technical point of view. A Knowledge Exchange Report. Available from: www.knowledge-exchange.info/datametrics. [Accessed on July 15 2013]

Hrynaszkiewicz, I.; Altman, D.G. (2009). Towards an agreement on best practice for publishing raw clinical trial data. *Trials*, vol. 10 (17). http://dx.doi.org/10.1186/1745-6215-10-17, PMid:19296844 PMCid:PMC2662833.

Moed, H.F. (2005). *Citation Analysis in Research Evaluation*. Dordrecht; Springer.

Piwowar, H.A.; Day, R.S.; Fridsma, D.B. (2007). Sharing detailed research data is associated with increased citation rate. *Plos One*, vol. 2 (3), e308. http://dx.doi.org/10.1371/journal.pone.0000308, PMid:17375194 PMCid:PMC1817752.

Rennolls, K. (1997). Science demands data sharing. *British Medical Journal*, vol. 315 (7106), 486.

Thomson Reuters (2012). Repository Evaluation, Selection, and Coverage Policies for the Data Citation Index within Thomson Reuters Web of Knowledge. Available from: http://wokinfo.com/media/pdf/DCI_selection_essay.pdf [Accessed on July 15 2013]

Torres-Salinas, D.; Robinson-García, N.; Cabezas-Clavijo, Á. (2012). Compartir los datos de investigación: introducción al data sharing. *El Profesional de la Información*, vol. 21 (2), 173-184. http://dx.doi.org/10.3145/epi.2012.mar.08

Torres-Salinas, D.; Robinson-García, N.; Campanario, J.M.; Delgado López-Cózar, E. (2013). Coverage, field specialization and impact of scientific publishers indexed in the 'Book Citation Index'. *Online Information Review*. [In Press].

Torres-Salinas, D.; Martín-Martín, A.; Fuente-Gutiérrez, E. (2013). An introduction to the coverage of the Data Citation Index (Thomson-Reuters): disciplines, document types and repositories. EC3 Working Papers (11), June 2013. http://arxiv.org/ftp/arxiv/papers/1306/1306.6584.pdf [Accessed on July 15 2013]

Vickers, A.J. (2006). Whose data set is it anyway? Sharing raw data from randomized trials. *Trials*, vol. 7 (15). http://dx.doi.org/10.1186/1745-6215-7-15, http://dx.doi.org/10.1186/1745-6215-7-30,PMid:17022818 PMCid:PMC1609186.

Wouters, P.; Schröder, P. (2003). *Promise and practice in data sharing: the public domain of digital research data*. The Netherlands: NIWI-KNAW.